\documentstyle[aps,prl,multicol,psfig]{revtex}\def\tc{yy}
\begin{document}
\title{Periodic-Orbit Theory of Anderson Localization on Graphs}
\author{Holger~Schanz$^{1}$ and Uzy Smilansky$^{2}$ }
\address{$^{1}$ Max-Planck-Institut f\"ur Str{\"o}mungsforschung,
Bunsenstr.\ 10, 37073 G{\"o}ttingen, Germany}
\address{$^{2}$
The Weizmann Institute of Science, Rehovot 76\,100, Israel}
\date{December 14, 1999}
\maketitle
\begin{abstract}
We present the first quantum system where Anderson localization is completely
described within periodic-orbit theory. The model is a quantum graph analogous
to an a-periodic Kronig-Penney model in one dimension. The {\it exact}
expression for the probability to return of an initially localized state is
computed in terms of classical trajectories. It saturates to a finite value
due to localization, while the diagonal approximation decays diffusively.  Our
theory is based on the identification of families of isometric orbits. The
coherent periodic-orbit sums within these families, and the summation over all
families are performed analytically using advanced combinatorial methods.
\end{abstract}
\if\tc\begin{multicols}{2}\fi 
Anderson localization is a genuine quantum phenomenon. So far, attempts to
study this effect within a semiclassical (periodic-orbit) theory seemed to be
doomed to fail from the outset: It is not clear whether the leading
semiclassical approximation for the amplitude associated with a single
classical orbit is sufficiently accurate. Even more seriously, there is no
method available to add coherently the contributions from the exponentially
large number of contributing orbits. Here, we address the second problem and
develop a method to perform the coherent periodic-orbit (PO) sums in a
standard model---a quantum graph analogous to the Kronig-Penney model in
1D---for which the PO theory is exact. For a list of references on the long
history of graph models see \cite{KS99}.

For investigating Anderson localization we consider the quantum return
probability (RP). It is defined as the mean probability that a wave packet
initially localized at a site is at the same site after a given time. We show
that the long-time RP approaches a positive constant, which proves that the
spectrum has a point-like component with normalizable eigenstates.  The
asymptotic RP is the {\it inverse participation ratio}, which is a standard
measure of the degree of localization. The RP can also be seen as 2-point form
factor of the {\it local} spectrum \cite{TDittrich}. As such, it belongs to
the class of quantities which can be expressed as double sums over PO's of the
underlying classical dynamics \cite{Ber85}. Because of the exponential
proliferation of the PO's in chaotic systems, the resulting sums are hard to
perform. Consequently, most semiclassical approaches to spectral two-point
correlations were restricted to the diagonal approximation where the
interference between different PO's is neglected
\cite{Ber85,TDittrich,JBS90,AIS93,Dit96}. While this method is very successful
for short-time correlations, it fails to reproduce long-time effects such as
Anderson localization which are due to quantum interferences. In \cite{ADD+93}
the universal long-time behavior of the form factor was related to universal
{\em classical action correlations} between PO's of a chaotic system. A deeper
understanding of how quantum universality is encoded in classical correlations
is highly desirable but still lacking, despite some recent progress
\cite{CPS98,Coh98}.  This context is our motivation for developing the first
PO theory of 1D Anderson localization, although the phenomenon as such is well
understood \cite{MT61,Ber74,T77,AALR79,KTI97}.

Quantum graphs exhibit both classical and quantum universal properties, which
qualify them as model systems in quantum chaos \cite{KS99}. The PO theory in
graphs is exact. Hence, graphs are well suited for the study of PO
correlations and their expected universal features. Recently, we reproduced
the complete form factor of the circular unitary ensemble of $2\times 2$
random matrices using a PO expansion in a simple quantum graph
\cite{SS99}. The new combinatorial tools developed there will be used to
compute the quantum RP, by extending a method due to Dyson \cite{Dys53} for
summing over orbits in a 1D topology.

\def\figI{
\begin{figure}[htb]
 \centerline{\psfig{figure=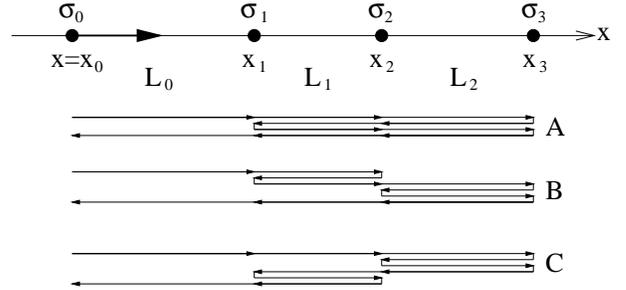,width=\figw}}
\vspace*{5mm}
\caption{\label{graph_inf}Top: Quantum chain graph with vertices (dots) and
bonds (line segments). The arrow shows the initial state used to compute the
RP. Bottom: The family $[m_{0},m_{1},m_{2}]=[1,2,2]$ of different but
isometric orbits returning after topological time $n=10$ to the vertex $0$.}
\end{figure}
}
\if\tc{\def\figw{8cm}\figI}\fi
We consider a quantum graph with a 1D chain topology and use the intuitive
notation as indicated in Fig.~1. The bond lengths $L_{j}=x_{j+1}-x_{j}$ are
disordered, i.~e.\ pairwise rationally independent. On a bond $j$ the general
solution of the 1D Schr\"odinger equation for wave number $k$ is
$\Psi_{j}(x)=a_{j,+}\exp(+{i}\,k[x-x_{j}])+a_{j,-}\exp(-{i}\,k[x-x_{j+1}])$.
The matching of solutions across a vertex is achieved in terms of a unitary
scattering matrix $\sigma_{j}(r_{j},t_{j})$ such that
\begin{equation}\label{smat}
\pmatrix{a_{j-1,-}\cr a_{j,+}}=
\pmatrix{t_{j} & r_{j} \cr r_{j} & t_{j}}\;
\pmatrix{{e}^{{i}\,k\,L_{j}}\ a_{j,-} \cr {e}^{{i}\,kL_{j-1}}\ a_{j-1,+}}
\end{equation}
The phase factors on the r.h.s. account for the free motion on the bonds.  The
matrices $\sigma_j$ will be assumed independent of $k$ and are parametrized as
$r_j=i\cos \eta_j, \ t_j = \sin\eta_j $.  A similar model was used e.~g.\ in
the analysis of an optical experiment demonstrating Anderson localization in
the transmission of light through a disordered stack of transparent mirrors
\cite{BK97}. The vertex-scattering matrices $\sigma_j$ could also be computed
by assuming $\delta$-potentials at $x=x_{j}$, as in the Kronig-Penney model
\cite{KP31}. In the following we shall consider two situations: (i) random
$\sigma$, where the $\eta_j$ are independent random variables distributed such
that the corresponding transmission and reflection probabilities $T_j =
|t_j|^2$, $R_j=|r_j|^2$ are uniformly distributed in the
interval $[0,1]$, and (ii) constant $\sigma$, where $T_j=T$, $R_{j}=R$
$\forall\ j$.

At fixed $k$ we consider the Hilbert space of coefficient vectors
${\bf a} \equiv a_{j,v}$ where $j$ goes over the vertices, and $v=\pm$.
We introduce a unitary operator
\begin{eqnarray}\label{bsm}
U_{j',v';j,v}(k)&=&{e}^{{i}\,kL_{j}}\,
(\delta_{+v',v}\delta_{j',j+v}\,t_{j+{v+1\over 2}}+
\nonumber\\
&& \phantom{{e}^{{i}\,kL_{j}}\,(}
\delta_{-v',v}\delta_{j',j}\,r_{j+{v+1\over 2}}
)\,.
\end{eqnarray}
The map defined by $U(k)$ describes the time evolution of a coefficient
vector. It is the natural object for the investigation of the graph \cite
{KS99}. $k$ is an eigenvalue of the graph Hamiltonian iff $1$ is in the
spectrum of $U(k)$, that is when all the matching conditions (\ref{smat}) can
be satisfied simultaneously ${\bf a} = U(k) {\bf a}$.  In terms of $U$ we can
characterize the degree of localization by calculating the averaged quantum RP
\begin{equation}
\label{qrp}
{\cal P}(n)=\left\langle |(U^{n})_{0,+;0,+}|^{2}\right\rangle
\end{equation}
as a function of the discrete topological time $n$ of a state which was
initially prepared as $a_{j,v}=\delta_{j,0}\delta_{v,+}$ (arrow in Fig.~1,
top). The average $\left\langle \dots\right\rangle $ in (\ref{qrp}) is over a
large $k$ interval, and for random $\sigma$ also over the ensemble of $\sigma
$ defined above.  Due to the structure of $U$, ${\cal P}(n)\ne 0$ only for
even $n=2m$.

The classical analogue of the quantum graph \cite{KS99} is a Markovian random
walk with vertex reflection and transmission probabilities which are equal to
the quantum mechanical ones defined above. A classical trajectory is encoded
by the sequence of traversed vertices $\{j_{\nu}\}$ which must be consistent
with the connectivity of the graph. In our case $v_{\nu}=j_{\nu+1}-j_{\nu}=\pm
1$. Given the initial vertex, a trajectory can also be identified uniquely by
the sequence $\{v_{\nu}\}$.  Because of the probabilistic nature of the
dynamics all trajectories are unstable.  In general, the representation of a
quantum evolution operator in terms of classical trajectories involves a
semiclassical approximation. Here it is {\em exact} and amounts simply to
expanding the matrix products in (\ref{qrp}). The result
\begin{equation}\label{po-expansion}
{\cal P}(n)=\left\langle \left|\sum_{\lambda} {\cal
A}_{\lambda}\exp({i}\,k{\cal L}_{\lambda})
\right|^{2}\right\rangle
\end{equation}
can be {\em interpreted} in terms of the classical trajectories introduced
above. $\lambda$ runs over all trajectories contributing to the RP
(\ref{qrp}), i.~e.\ all sequences $\{v_{\nu}\}$ ($\nu=0,\dots,n-1$) with
$\sum_{\nu}v_{\nu}=0$. The traversed vertices are
$j_{\nu}=\sum_{\nu'\le\nu}v_{\nu'}$.  For graphs, the concepts of returning
trajectories and PO's coincide, hence (\ref{po-expansion}) is a PO sum. The
length of an orbit is ${\cal L}_{\lambda}=\sum_{\nu=0}^{n-1}L_{j_{\nu}}$, such
that $k{\cal L}_{\lambda}$ is the corresponding action in units of
$\hbar$. The amplitude ${\cal A}_{\lambda}$ is the product of the transmission
and reflection amplitudes accumulated along the orbit ${\cal
A}_{\lambda}=\prod_{\nu=1}^{n}{\cal A}_{\lambda,\nu}$ with ${\cal
A}_{\lambda,\nu}=t_{j_{\nu}}$ if $v_{\nu+1}=v_{\nu}$ and ${\cal
A}_{\lambda,\nu}=r_{j_{\nu}}$ otherwise. It was shown in \cite{KS99} that
$|{\cal A}_{\lambda}|$ plays the r\^ole of the stability amplitude of the
orbit, while the phase of ${\cal A}_{\lambda}$ is equivalent to the Maslov
index.

\def\figII{
\begin{figure}[htb]
 \centerline{\psfig{figure=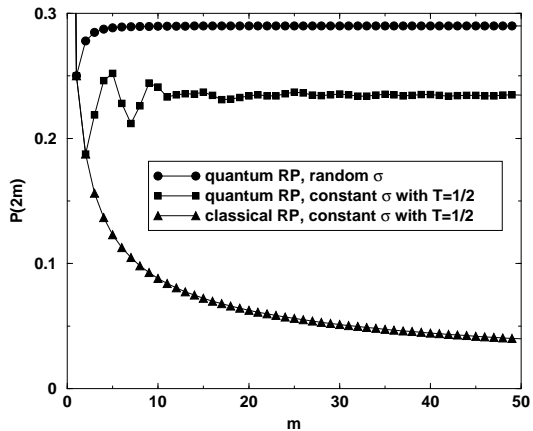,width=\figw}}
\caption{\label{rp} Exact quantum and classical return probability
${\cal P}(2m)$ as given by Eqs.~(\protect\ref{crp}), (\protect\ref{rp2s}). }
\end{figure}
}
\if\tc{\def\figw{7cm}\figII}\fi
Expanding $\left|\dots\right|^{2}$ in (\ref{po-expansion}) we obtain a double
sum over PO's of the type studied in \cite{Ber85}.  For short time, the
interference terms pertaining to two different PO's which are not related by
an exact symmetry can be neglected due to the averaging applied. For constant
$\sigma$ with $T=R=1/2$ this {\em diagonal approximation} simply amounts to
counting the number of classical PO's with period $n=2m$ since all weights
$|{\cal A}_{\lambda}|^2=2^{-n}$ coincide. Any such PO is represented by a code
word $\{v_{\nu}\}$ containing $m$ letters $+$ and $-$, respectively. According
to the initial condition $v_{1}=+$. Hence, each PO $\lambda$ in
(\ref{po-expansion}) corresponds to a way of selecting from the remaining
$n-1$ time steps those $m$ with negative velocity, and we have
\begin{equation}\label{crp}
{\cal P}_{\rm cl}(n=2m)={1\over 2^{n}}{n-1\choose n/2}\approx{1\over
\sqrt{2n\pi}}\qquad({n\to\infty})
\end{equation}
(triangles in Fig.~2). It is well known that the diagonal approximation yields
the {\em classical} RP \cite{TDittrich}. And indeed, for long time (\ref{crp})
shows the expected classical diffusion. The decay of the classical RP to $0$
corresponds to the obvious fact that there is no localization in the classical
analogue of our model.

In the following we will show that a constructive interference between PO's
with different number of reflections but equal lengths leads to a finite
saturation value of the exact RP and consequently to localization
\cite{discordance}.  When (\ref{qrp}) is expanded into a double sum, only
pairs of orbits with equal lengths survive the $k$ average.  Hence, all
relevant interferences are confined to families of {\em isometric}
orbits. Suppose a PO $\lambda$ returning to bond $(0,+)$ after $2m$ collisions
with vertices traverses the bonds $(j,+)$ and $(j,-)$ $m_{j}$ times,
respectively ($\sum_{j}m_{j}=m$). The length of this PO is ${\cal
L}_{\lambda}=2\sum_{j}m_{j}L_{j}$. Thus, for rationally independent bond
lengths $L_{j}$ a family of isometric orbits contains all orbits sharing the
set ${\cal L}=[m_{0},m_{\pm 1},\dots]$. In contrast to the diagonal
approximation, these are not only symmetry-related orbits. A simple example is
shown in Fig.~1 (bottom), but with increasing orbit length families can
contain more and more PO's. Taking into account orbit families
Eq.~(\ref{po-expansion}) becomes
\begin{eqnarray}\label{lc-expansion}
{\cal P}(n=2m)=\sum_{{\cal L}\in{{\cal F}_{n}}}\left|
\sum_{\lambda\in{\cal L}}{\cal A}_{\lambda}\right|^2
=\sum_{{\cal L}\in{{\cal F}_{n}}}{\cal P}({\cal L})\,.
\end{eqnarray}
The outer sum is over the set ${\cal F}_{n}$ of families, while the inner one
is a {\it coherent} sum over the orbits belonging to a given family. The phase
and amplitude of each PO depend on its itinerary. The phase equals the parity
of half the number of reflections. Had one assumed that these phases are
randomly distributed within a family, (\ref {lc-expansion}) would again reduce
to the diagonal approximation \cite{long}. In contrast, the exact result
derived below from Eq.~(\ref{lc-expansion}) is shown in Fig.~2. For both,
constant and random $\sigma$, the RP saturates to a finite value indicating
localization.  Hence, quantum localization is due to a delicate and systematic
imbalance between positive and negative terms within families. This can be
regarded as a classical correlation (deviation from a random distribution) of
phases of PO's. The saturation value, i.~e.\ the inverse participation ratio
is expected to be inversely proportional to the localization length and the
classical diffusion constant, see e.~g.\ \cite{Dit96} and
refs. therein. Indeed we find for constant $\sigma$ and a diffusion constant
$D=T/(1-T)\gg 1$ the relation $\lim_{m\to\infty}{\cal P}(2m)=(D\pi)^{-1}$
\cite{long}.

In the sequel we outline the analytical evaluation of
Eq.~(\ref{lc-expansion}), deferring a detailed expos\'e to a subsequent
publication \cite{long}. A sum with a similar structure was calculated by
Dyson \cite{Dys53} to study a disordered chain of harmonic oscillators. This
system is analogous to our model, but Dyson computed a different
quantity---the density of states. It is essentially different from the RP
because the latter is a two-point correlation function.  Dyson computed the
traces of powers of a {\it Hermitian} matrix, and did not have to keep track
of the phases encountered in the computation of diagonal elements of powers of
a {\it unitary} matrix.  Consequently the combinatorial arguments needed to
evaluate (\ref{lc-expansion})---though similar in spirit to Dyson's
approach---are quite different from \cite{Dys53}.

To start, we consider a one-sided graph with $j\ge 0$, and $R_{0}=1$, and
perform the coherent sum over all orbits $\lambda$ in a given family.  To keep
track of the precise number of reflections for each orbit, we choose to
specify the orbit by the segments with positive velocity $v_{\nu}=+$. In the
schematic representation of Fig.~1 (bottom) these are the arrows which {\em
point to the right}. When the left pointing arrows are deleted from the bonds
$1$ and $2$ in Fig.~1 (bottom) one obtains the arrow structures displayed in
Fig.~3. The {\it vertical} displacement of the arrows provides the information
about the time ordering which is necessary to reconstruct the complete orbit
from the right pointing arrows. The lower an arrow, the later it appears in
the trajectory. The number of arrows to the left (right) of the vertex $j$ is
$m_{j-1}$ ($m_{j}$). An orbit can be completely specified by prescribing, at
all vertices, in which order the adjacent arrows are traversed. Note that this
local time order can be chosen independently for each vertex. As a
consequence, the factorization
\begin{equation}\label{fac}
{\cal P}([m_{0},\dots,m_{b-1}])={\cal V}_{b}(m_{b-1})\prod_{j=1}^{b-1}{\cal
V}_{j}(m_{j},m
_{j-1})
\end{equation}
results. $b$ denotes the rightmost vertex along the orbit, i.~e.\ $m_{j}=0$
$\forall j\ge b$. The local contribution ${\cal V}_{j}(m_{j},m_{j-1})$ from a
vertex will be determined in the following.

\def\figIII{
\begin{figure}[htb]
 \centerline{\psfig{figure=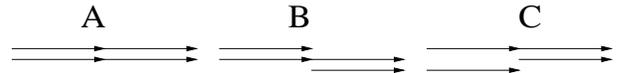,width=\figw}}
\vspace*{5mm} \caption{\label{local-order}This figure corresponds to vertex 2
in Fig.~1, which is approached twice from the left and from the right.  The
right pointing arrows can be arranged in 3 distinct ways, which correspond to
the 3 decompositions of the number 2 into 2 non-negative integers
2=1+1=0+2=2+0 (left to right).}
\end{figure}
}
\if\tc{\def\figw{8cm}\figIII}\fi
Each local time order at vertex $j$ corresponds to one of
${m_{j}+m_{j-1}-1\choose m_{j-1}-1}$ possibilities to distribute $m_{j}$
identical objects over $m_{j-1}$ sites, or---in other words---to decompose
$m_{j}$ into $m_{j-1}$ non-negative summands (Fig.~3). The number of
transmissions to the right at vertex $j$ is given by the number of non-zero
terms in this decomposition. We can express ${\cal V}_{j}(m_{j},m_{j-1})$ as a
sum taking into account the number $\nu$ of such transmissions, together with
the associated amplitude.  We obtain
\begin{equation}\label{kravtchouk}
{\cal V}_{j}= \left|\sum_{\nu}
{m_{j-1}\choose\nu}{m_{j}-1\choose\nu-1}t_{j}^{2\nu}\,r
_{j}^{m_{j}+m_{j-1}-2\nu} \right|^{2}\,.
\end{equation}
The standard definition of binomial coefficients ensures that the sum can be
taken over all integers $\nu$. The power of $t_{j}$ in the expansion
(\ref{kravtchouk}) is $2\nu$ since the number of transmissions to the left and
to the right is the same. ${m_{j}-1\choose\nu-1}$ is the number of
decompositions of $m_{j}$ into $\nu$ positive summands and
${m_{j-1}\choose\nu}$ corresponds to selecting the $\nu$ non-zero terms from
all $m_{j-1}$ summands. The sum in (\ref{kravtchouk}) is a special
Kravtchouk polynomial---a well-known object in combinatorics
\cite{NSU91,SS99}.

With (\ref{fac}), (\ref{kravtchouk}) we can perform the summation over all
orbits $\lambda$ of a given family in (\ref{lc-expansion}). However, there are
$\sum_{b=1}^{m}{m-1\choose b-1}=2^{m-1}$ decompositions of an integer $m$ into
positive summands $m_{j}$, i.~e.\ the number of families grows exponentially
with time. To sum over all families we derive a recursion relation which
dramatically reduces the number of terms needed. Grouping in
(\ref{lc-expansion}) all terms according to the number of traversals of the
first two bonds $m_{0}$ and $m_{1}$, i.~e. defining
$B_m(m_{0},m_{1})=\sum_{m_{2},\dots} {\cal P}([m_{0},m_{1},m_{2},\dots])$, we
find $B_{m}(s+1,t)={\cal V}(t,s+1)/{\cal V}(t,s)\,B_{m-1}(s,t)$ for $s\ge0$
and $B_{m}(1,t)={\cal V}(t,1)\sum_{s=1}^{m-2}B_{m-1}(t,s)$. Summing over the
second argument of $B_{m}(s,t)$ we obtain the combined contribution $V_{m}(s)$
from all orbits with period $2m$ which traverse the initial bond exactly
$s$ times before finally returning to it. The corresponding recursion relation
is
\begin{equation}\label{recursion}
V_{m}(t)=\sum_{s=1}^{m-t}{\cal V}(
s,t)V_{m-t}(s)\qquad (1\le t<m)\,.
\end{equation}
The recursion is initialized using the elements $V_{m}(m)$, which are due to a
single PO: the orbit bouncing $m$ times between the vertices $j=0$ and $j=1$.
The RP for the one-sided graph is now ${\cal P}(2m)=\sum_{s=1}^{m}V_m(s)$.

For the unrestricted graph a returning trajectory can be composed of simple
loops to the right and to the left from the initial bond. Both groups can be
described by the results for the one-sided problem. Using similar arguments as
in the derivation of (\ref{kravtchouk}) we find
\begin{equation}\label{rp2s}
{\cal P}(2m)=\sum_{m_{r}=1}^{m}
\sum_{s=1}^{m_{r}}V_{m-m_{r}+s}(s)\,V_{m_{r}}(s)\,.
\end{equation}
We were able to solve the recursion (\ref{rp2s}) analytically for random
$\sigma$.  After expanding $|\dots|^2$ in (\ref{kravtchouk}) we get
\def\ls{
 {\cal V}_{j}^{\rm (av)}(m_{j},m_{j-1})
}
\def\rsI{
 {1\over 2}\int_{0}^{\pi/2}\!\!\!\!\sin 2\eta_j {\rm d}\eta_{j}\, {\cal
 V}_{j}(m_{j},m_{j-1},\eta_{j})
}
\def\rsIIa{
 {m_{j}^{2}\over m_{j-1}+m_{j}+1}\sum_{\nu,\nu'}{(-1)^{\nu+\nu'}\over \nu\nu'}
 {m_{j-1}+m_{j}\choose \nu+\nu'}^{-1} 
}
\def\rsIIb{
 {m_{j-1}-1\choose \nu-1}{m_{j}- 1\choose \nu-1} {m_{j-1}-1\choose
 \nu'-1}{m_{j}-1\choose \nu'-1}
}
\def\rsIII{
 {2\,m_{j-1}^{2}\over
 (m_{j-1}+m_{j}-1)(m_{j-1}+m_{j})(m_{j-1}+m_{j}+1)}
}
\if\tc
\begin{eqnarray}
&&\
\ls = \rsI \nonumber\\
&=& \rsIIa \nonumber\\
&&\times \rsIIb \nonumber
\\
&=& \rsIII \,.
\label{avvp}
\end{eqnarray}
\else
\begin{eqnarray}
\ls
&=& \rsI \nonumber\\
&=& \rsIIa \nonumber\\
&&\times \rsIIb \nonumber
\\
&=& \rsIII \,.
\label{avvp}
\end{eqnarray}
\fi The last equality follows from an identity which we proved previously
\cite{SS99} using recent developments in combinatorial theory \cite{a=b}.  For
the outmost vertex on an orbit (\ref{avvp}) simplifies to ${\cal P}_{b}^{\rm
(av)}(m_{b-1})=1/(m_{b-1}+1)$. Consequently we can initialize the recursion
(\ref{recursion}) by $V^{\rm(av)}_{m}(m)=1/(m+1)$, which results in
$V^{\rm(av)}_{m}(s)=s/(m^{2}+m)$. Summing with respect to $s$ we find that the
RP of the one-sided graph is constant ${\cal P}_{\rm os}(2m)=1/2$ for $m>0$.
For the unrestricted graph and $m\to\infty$ the double sum in (\ref{rp2s}) can
be approximated by a double integral yielding $\lim_{m\to\infty}{\cal
P}(2m)=\pi^{2}/3-3$ which is indeed the saturation value of the top curve in
Fig.~2.

In summary, we have shown that Anderson localization can be reproduced from PO
theory only when classical correlations are properly taken into account. Here,
these correlations show up as exact isometries of families of PO's, and the
non-random distribution of the phases within a family. The failure of the
diagonal approximation in a model where PO theory is exact, identifies the
neglect of action correlations as the primary source of errors in
semiclassical theories of localization and related problems.

Support by the Minerva Center for Nonlinear Physics and the Israel Science
Foundation is acknowledged. We thank U.~Gavish for valuable comments.

\if\tc\else
\def\figw{18cm
}
\figI
\figII
\figIII
\fi
\if\tc \end{multicols}\fi

\begin{thebibliography}{10}
\bibitem{discordance} This constructive interference is complementary to the
``discordance'' (destructive interference) between non-returning rays
(trajectories) which was discussed in \cite{BK97}.
\end{thebibliography}
\end{document}